# Hydrostatic Level Sensors as High Precision Ground Motion Instrumentation for Tevatron and Other Energy Frontier Accelerators

*James Volk[a], Sten Hansen[a], Todd Johnson[a], Hans Jostlein[a], Terry Kiper[a], Vladimir Shiltsev[a], Andrei Chupyra[b], Mikhail Kondaurov[b], Anatoly Medvedko[b], Vasily Parkhomchuk[b], Shavkat Singatulin[b], Larry Stetler[c], Jason Van Beek[c], Dante Fratta[d], Josh Roberts[d], Herb Wang[d]*

[a]*Fermi National Accelerator Laboratory, PO Box 500, Batavia, IL 60510, USA*

[b]*Budker Institute of Nuclear Physics, Novosibirsk, 630090, Russia*

[c]*South Dakota Schools of Mines and Technology*

[d]*University of Wisconsin-Madison, Madison, WI 53706, USA*

*E-mail*: volk@fnal.gov

ABSTRACT: Particle accelerators pushed the limits of our knowledge in search of the answers to most fundamental questions about micro-world and our Universe. In these pursuits, accelerators progressed to higher and higher energies and particle beam intensities as well as increasingly smaller and smaller beam sizes. As the result, modern existing and planned energy frontier accelerators demand very tight tolerances on alignment and stability of their elements: magnets, accelerating cavities, vacuum chambers, etc. In this article we describe the instruments developed for and used in such accelerators as Fermilab's Tevatron (FNAL, Batavia, IL USA) and for the studies toward an International Linear Collider (ILC). The instrumentation includes Hydrostatic Level Sensors (HLS) for very low frequency measurements. We present design features of the sensors, outline their technical parameters, describe test and calibration procedures and discuss different regimes of operation. Experimental results of the ground motion measurements with these detectors will be presented in subsequent paper.

## Contents

## 1. Introduction: The Need for Hydrostatic Level Systems

Present energy frontier accelerators have beam spot sizes in the interaction regions on the order of tens of micrometers, e.g. some 30 micrometer in the Tevatron 1.96 TeV center of mass (com) energy proton-antiproton collider, and 20 micrometer in the CERN's Large Hardron Collider (LHC) at 7 TeV com. Given that these accelerators are several kilometres understanding and compensating for motion of individual components due to natural ground motion and cultural sources of motion is imperative as element misalignments lead to off-centered collisions at the interaction points, as well as several other unwanted effects. General consideration of these effects and tolerances can be found in Reference [1] and a recent review [2], as well as numerous references therein. Proposed future accelerators such as the International Linear e+e- Collider (ILC) [3] or the Compact Linear Collider (CLIC) [4] will have spot sizes on the order of nanometres and their performance will be very dependent on minuscule ground motion. Very schematically, from accelerator prospective, the ground motion can be divided into two regions; slow (i.e., often at frequencies below 1 Hz) and fast (frequencies greater than 1Hz). Hydrostatic Level Systems (HLS) are ideal for the study of slow ground motion. Future accelerators will be dependent on HLS to monitor ground motion and use that information in feedback and feed forward systems to adjust machine components to compensate for movement.

## 2. Fundamentals of HLS

The fundamental principle of hydrostatic water levels systems [5] is that any fluid seeks its own level. Given two pools connected by a pipe or tube the fluid level in each pool will be at the same absolute elevation. For small systems the absolute elevation can be relative to some reference such as mean sea level, for longer systems the curvature of the earth and other gravitational effects need to be considered. Given a reference point such as the bottom or top of the housing the relative level of the fluid will vary as the sensor is moved up or down with respect to all the other sensors. For large systems the differential gravity effects do need to be considered. While any fluid can be used water has the advantage of a low viscosity allowing for movement between pools and it presents no environmental hazards in case of spillage, distilled water is easily obtained and with sufficient care there will be no growth of bacteria or biological systems in the pools or tubing. There exists the problem of evaporation of the water from the surface of the pools. Even closed systems have some small evaporation issues. Some systems use fine oil films such as silicon diffusion pump oil on the top of the water in each pool to reduce evaporation.

### 2.1 Water levels

There are many types of water levels measuring systems. These have been extensively reported in the International Workshop on Accelerator Alignment [6] and in geological literature [7] and [8].

### 2.2 Plumbing

There are two types of plumbing systems in use. The single pipe or half filled system, and the two-pipe or fully filled system. It is important that both the water volumes and the air volumes are connected. The single pipe communicates both water and air to each of the pools in one pipe. The two-pipe system has separate pipes one fully filled with water the other full filled with air.

The two pipe system is more affected by temperature variations depending on the vertical column of water. As the temperature along the system changes the volume of water will expand and contract causing spurious reading. This can be removed by careful monitoring of the temperature along the entire system and correcting the data based on temperatures. The two-pipe system has the ability to avoid obstacles by changes in elevation of the water pipe. This is important for installation around existing equipment.

The single pipe system is not as affected by temperature and pressure changes but does require that the entire piping be level throughout the system. The diameter of the pipe should be sufficiently large to avoid air blocks or bubbles. If the pipe varies in elevation by more that its diameter air bubbles or water blockage can occur. This problem has caused inaccurate readings in systems due to lack of communication of the fluid used.

### 2.3 Types of Sensors

To measure the level of the fluid relative to a fixed point several means can be used such as; capacitance of the air above the fluid level that can be converted into a distance, ultra sound to measure from the depth from below or the air above the fluid, and laser time of flight either through the fluid or the air above the fluid to determine fluid level [6].

### 2.3.1 Capacitive Sensors

In the water pool there are two medium, air and water. Figure 1 shows a simplified view of the pool. Where C stands for the capacitance, ε the relative permittivity of the medium, and $D_1$ the distance between the sensor surface and the water and the $D_2$ the depth of the water.

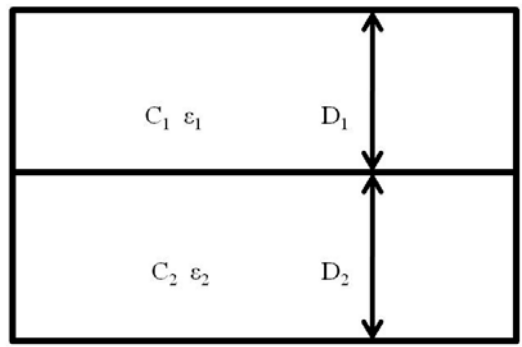

**Figure 1:** Schematic of HLS pool showing capacitances and distances.

The total capacitance is given by

$$1/C = 1/C_1 + 1/C_2$$

and

$$C_1 = \varepsilon_1 A_1/D_1 \quad C_2 = \varepsilon_2 A_2/D_2$$

where $A_1$ and $A_2$ are constants defined by the geometry of the surface area. Substituting the equations for each capacitance into the first equation and solving for $D_1$ gives;

$$D_1 = A_1 E_1/C(1-k) + Dk/(1-k)$$

Where $D = D_1 + D_2$ is the total distance and $k = A_1\varepsilon_1/A_2\varepsilon_2 \quad k<<1$

The values of $\varepsilon_2$ for water is 81 times that of air $\varepsilon_1$ so the assumption that k is much less than one is valid. Measuring the changing capacitance C gives the water level $D_1$ according to;

$$D_1 = \alpha/C + \text{offsets}$$

Where the offsets and scale $\alpha$ are defined in the calibration process. The actual capacitance for a 5 mm gap has been measured to be close to 0.5 pico-farads.

### 2.3.2 Ultra Sonic Sensors

Another method to determine water levels employs megahertz ultrasonic transponders. A transponder emits a short sonic pulse that reflects from the surface of the water. The time of flight is digitized and converted into a distance using the velocity of sound in water. Fixed reflecting surfaces are incorporated in the vessel to allow for real time calculation of the velocity of sound thereby removing the temperature dependence issue. The top of the pillar is machined to accept a survey ball thereby allowing linkage to and outside survey network.

## 3. Budker type Sensors

Fermilab has had a long relationship with the Budker Institute of Nuclear Physics (BINP) in Novosibirsk, Russia. Since 2001 the Budker Institute designed, developed and produces three types of HLS sensors, two of these are capacitive sensors named SAS and SAS-E and the other is an ultra sound named ULSE. The SAS and SAS-E sensors use similar electronics and compatible pools the main difference is that the SAS data and power is supplied via a daisy chain cable system while the SAS-E sensors uses Power over the Ethernet (PoE) switch for power and data acquisition. Each SAS-E sensor is connected directly to the PoE switch this allows for a radial configuration of sensors. The advantage of the SAS-E sensors is its ability to set individual parameters for each sensor via the switch when the sensors are in use. The disadvantage is the sensitivity to radiation that can cause bits to flip in the digital electronics and parameters, such as the IP or MAC address to change at unexpected times.

### 3.1 Budker Capacitive Sensors

The sensors have two parts: the pool and the electronics. The shells of the sensors are made of 304 stainless steel. There is a port in the pool section for insertion of a thermometer. Figure 2 shows a typical sensor split in two the face of the upper part is the capacitive pick up. The data and power are supplied by a daisy chain cable each sensor has a unique serial number that allows for logging of data and calibration parameters unique to each sensor to be used in the data reduction. There is also a thermometer to monitor the temperature of the stainless steel pool and hence the water temperature.

## 3.2 Calibration of SAS and SASE type sensors

### 3.2.1 Calibration done at Budker

At the Budker Institute there is a movable stage that is used to calibrate each sensor. The sensor is set on the stage and data are taken at 5, 6, 7, 8, 9 and 10 mm from the electrode surface. A polynomial fit is done to the data to generate the calibration values. By this method the non linearity in the measurement of the capacitance are removed. The non linearity's in the sensor response is caused by the relative sizes of the active and guard regions of the capacitors. The calibration file is prepared and is read in by the system interface and stored in the on line PC for correction of measured distances.

### 3.2.2 Field Calibration

From time to time it is necessary to open the electronics package for repair. As it is not possible to return the sensors to Budker Institute or detail calibration a simple calibration procedure using two stainless steel disks (one giving a 5 mm air gap and the other 10 mm air gap relative to the electrode head) is used. A LABVIEW based program runs one sensor at a time. The 5-mm thick disk is placed into a standard pool and the sensor with no O-ring is placed on top. Ten measurements of the distance from the face of the sensor to the disk are made. The 5-mm disk is then removed from the pool and the 10-mm disk is inserted. Another ten measurements are made of this distance. A least squares fit is made to the data resulting in the necessary calibration values. These do not take into account the non lineararity of the sensor. The data are then added by hand to the master calibration file.

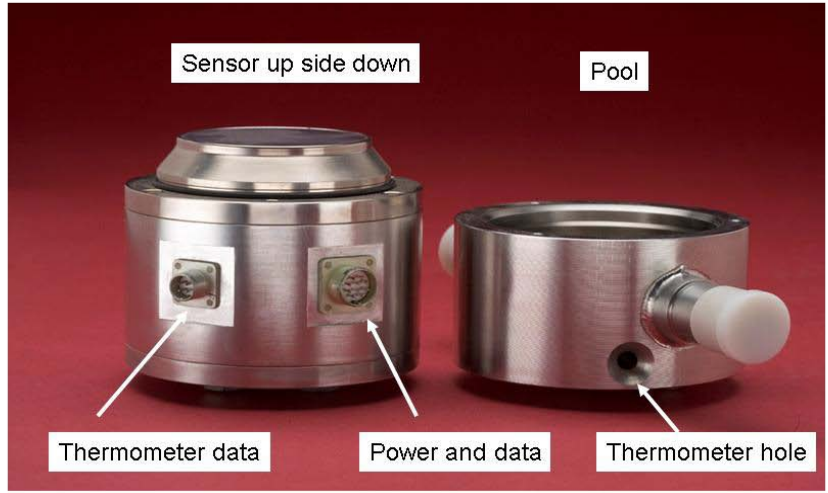

**Figure 2:** SAS type HLS sensor from Budker INP (Novosibirsk, Russia)

### 3.3 Budker Ultrasonic Sensors

To achieve higher resolution than the capacitive sensors the Budker Institute developed a system that uses ultra sound to measure the level of the water in a pool. The ultrasonic transponder manufactured by General Electric (type H10KB5) has a frequency range of 5 to 10 MHz with typical response of 7 MHz, and a relative pulse echo sensitivity of -48 to -38 dB. The transponder is mounted in the bottom of the pool with a precision machined stainless steel post above figure 3. There is a step machined in the post. This step is measured by a CMM to determine the exact spacing. When an ultrasonic pulse is emitted there are three reflections detected.

Lstep is the measured distance between the two steps in the stainless steel pieces. It is equal to D1 distance in figure 3. It is put in as Us_L0 in the settings page. Us_L0 is equal to twice Lstep the velocity of sound is measured to be;

$$V_s = \frac{2 \cdot L_{step}}{R_2 - R_1} = \frac{2 \cdot D_1}{R_2 - R_1} \quad (1)$$

It should be noted that R1 and R2 are total transit times as measured by the ultra sonic transponder. That is the time it takes the sound wave to go from the transponder to the surface and return to the transponder.

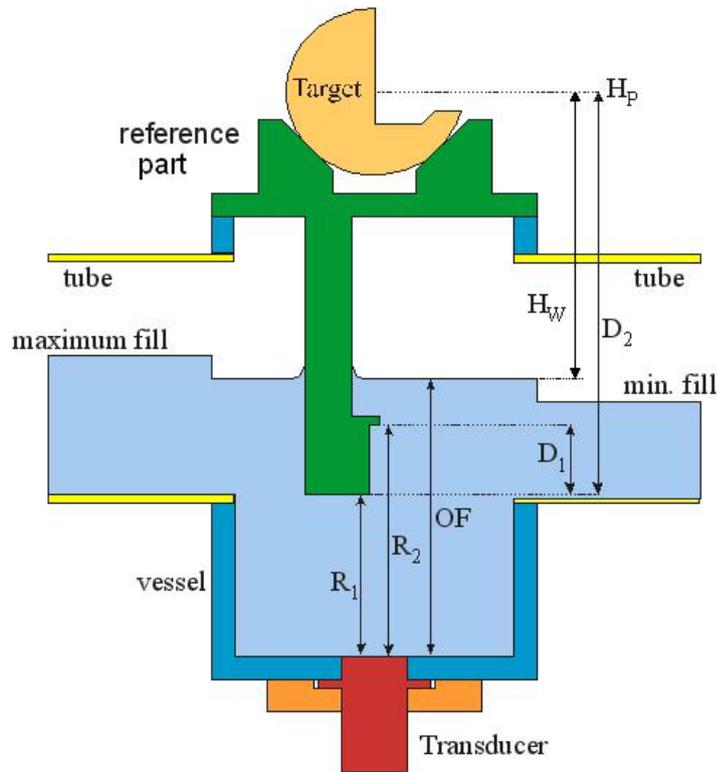

**Figure 3**: Side view of Budker Ultra Sonic Sensor showing pillar and steps

The distance from the bottom of the pipe to the top of the water is given by

$$\text{Top of water} = V_s * (OF-R_1) / 2 \quad (2)$$

The factor of two is needed to account for the double transit of the sound pulse. The distance Hp is also known by construction and external measurement. This is the distance between the center line of the target and bottom of the post in the sensor. This is put into the setting file as Us_L1. Hw is then calculated as

$$H_w = H_p - V_s * (OF-R_1)/2 - Us\_L2 \quad (3)$$

where Us_L2 is a constant that relates target to the top of the water pipe it is in the settings file. This constant can be changed to suit the particular setup of water levels. Equation 3 can be written as

$$H_w = H_p - 2D_1/(R_2-R_1) * (OF-R_1)/2 - Us\_L2 \quad (4)$$

and simplifying

$$Hw = Hp - D1/ (R2-R1) * (OF-R1) - Us\_L2 \quad (5)$$

Or in terms of the input to the program

$$Level = Us\_L1 – Us\_L0 *(OF -R1)/ (R2-R1) - Us\_L2 \quad (6)$$

where Us_L0, Us_L1, and US_L2 are distances in nano meters and OF, R1, R2 are times in nano seconds.

The electronics for the sensor are located in a separate box from the pool and transducer. The electronics can be up to 2 meters from the pool and transponder. This is to eliminate the issue of radiation damage to the analogue and digital electronics. Figure 4 shows a ULSE sensor and electronics. The functional circuit diagram of ULSE electronics is presented on the Fig.5. The electronics includes Transmitter/Receiver, Comparator, Time Digital Converter (TDC), Flash microcontroller of MSC1210 Type of Texas Instruments Corp., System clock oscillator, XPort, Pover controller, DC/DC converters and Transformer. All the electronics are operated by the microcontroller. The latter is controlled by the commands of Operator Board computer via Power over Ethernet interface. The microcontroller's algorithm of autonomous operation is distributed inside its internal flash memory.

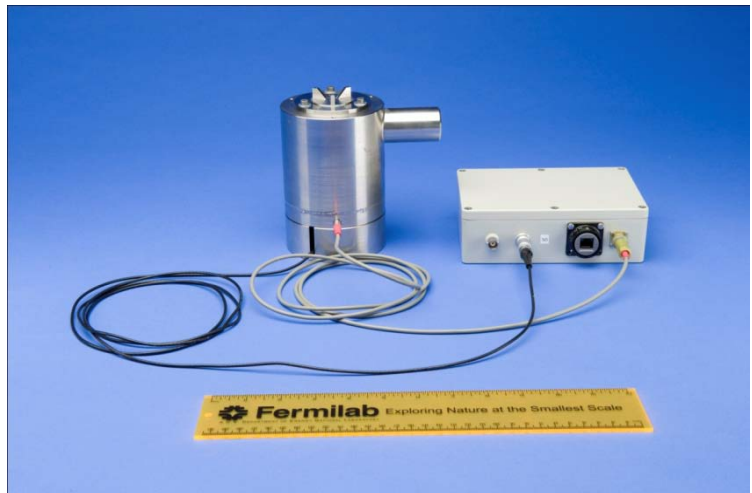
**Figure 4:** Ultrasonic sensor and electronics

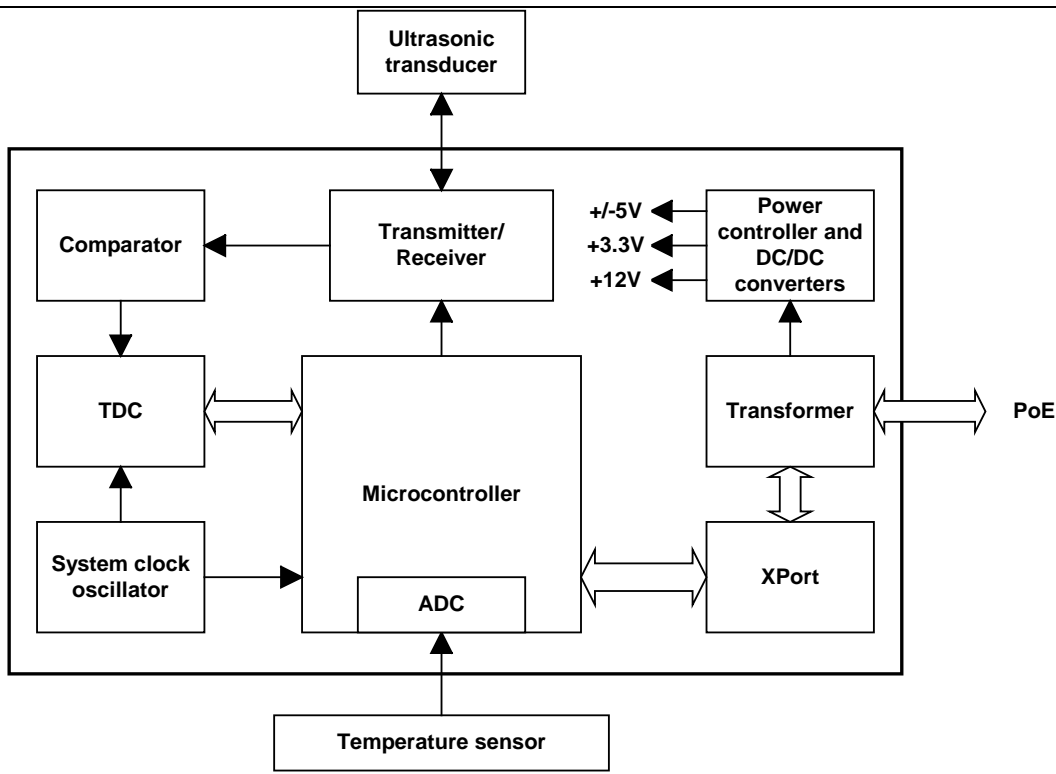

**Figure 5:** Block diagram of the ULSE type electronics.

On power up the microcontroller initializes the electronics from a program in its internal memory. The PC runs the measurement cycle by sending the start pulse for the transmitter and TDC. The transmitter pulses the transducer in the pool and the receiver acquires the reflected pulses and sends them to the zero level comparator. There is a gate based on the predicted timing of the reflected pulses for the first positive pulse after the first negative pulse of sufficient amplitude the comparator generates a stop signal for the TDC see figure 6.

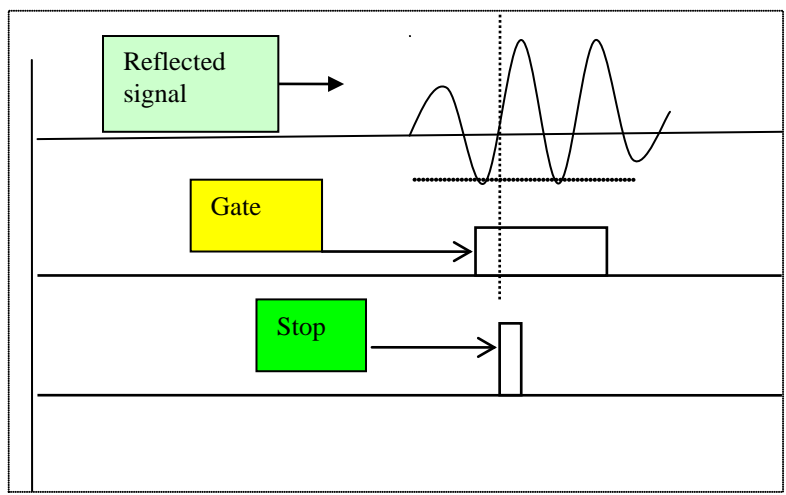

**Figure 6**: Timing of signals for ULSE type sensors

TDC measures time intervals between the start pulse and the pulses coming from the Comparator. The microcontroller gets digital codes of the measured time intervals from TDC and transforms them to the distance digital codes for the next transmission to PC computer. The measurement resolution is ± 0.15 µm. The microcontroller also can measure temperature of ULSE vessel and accordingly temperature of water inside the vessel with help of temperature sensor and onboard ADC. Temperature resolution is 0.1 $^{o}$C. The work of the microcontroller and TDC is synchronized by System clock quartz oscillator.

XPort is used as server for 10BASE-T/100BASE-TX Ethernet connection. The power supply controller and DC-DC protects the electronics from spikes. The input voltage of the converters can be as high as 48V. The output voltages are +12V, ±5V and +3.3V. The transformer is used for galvanic insulation between external electric circuits and the sensor electronics.

### 3.4 Data Acquisition process

The capacitive measurements are converted into a frequency and then digitized locally in each sensor. This avoids the problem of noise and signal loss in analogue signals sent long distances. This signal is then converted into a digital pulse and sent to the PC over a serial RS485 link (National Instruments PCI 232/485 card) for SAS sensors or Ethernet LAN under control of TCP/IP network protocol for SAS-E and ULSE. The SAS-E and ULSE sensors use Power over the Ethernet (PoE) to supply power to the sensors and transmit the digital signal to the data acquisition computer. The PoE switch has the ability to turn off any sensor in the system, and to adjust individual parameters of each sensor via the Ethernet connection.

To make the data acquisition programs simple the process is split into three parts that run under the Windows operating system. There is a service program called HlsService that runs in the background of the machine and is responsible for data acquisition and logging data. There is the HlsInterface program that the users interact with to set parameters for the system, and inspect the data in real time. There is also a program LGMvisio that allows for off line data analysis. This program can plot sensors or make ASCII files that can be analyzed by programs such as MS EXCEL or ORIGIN or used as an input file to C++ programs.

The HlsService program generates three files these have extensions .hdr, .err, and .lev. The .hdr file contains the start time and number of sensors being read out. The .err file contains any error messages recorded during data acquisition. The .lev file contains the data. The data acquisition process is stopped at the end of every month. At that time a system reset is sent and a new set of data files are started.

### 3.5 Stability of sensors
#### 3.5.1 Water Stability

The volume expansion of water is a nonlinear as a function of temperature [7]. From 0 $^{o}$C to 4 $^{o}$C the volume contracts above 4 $^{o}$C the volume increases slowly at first and then more rapidly. This can lead to nonlinear behavior in the level measurements particularly for the two pipe fully filled systems due to the non linear expansion of the water in the tubes. Also the volume is affected by changes in atmospheric pressure if there are vertical columns of water. The single pipe half filled system is not as sensitive to these changes. Temperature measurements at each sensor allow for correction of the data to remove these errors. Given the know coefficient of expansion for water at $209 \times 10^{-6}/°C$ the changes can be calculated. Systems that test stability are kept small to further reduce these effects.

There is also a slow evaporation of water from the systems. The SAS and SASE type capacitive sensors have heaters on the ceramic pickup to prevent condensation from forming. For systems of 4 or more sensors a loss of 1 micro meter of water level per day it typical. A thin layer of fine oil can be put on the water in the pools to help reduce the evaporation problems.

### 3.5.2 Electronic Stability and Overall Stability

To understand measurements at the micro meter level the time and temperature stability of the sensors must be fully understood. Studies on the SASE and ULSE sensors were preformed in 2010. One influence is temperature effects on the electronics. For SASE type sensors a fixed disk was inserted into the water pool, the sensors were placed in an insulated box and heated to 30 C and let cool very slowly. The distances were measured and plotted as a function of temperature figure 7a shows a typical plot. The curve can be fit and correction of electronics drift due to temperature can be made. A pair of the probes set side-by-side shows the differential noise level of $\sigma^2 = (0.09 \mu m)^2 + 1.252 \cdot 10^{-7} \ \mu m^2/s \cdot T$ as shown in Fig.7b (more details can be found in reference [1])

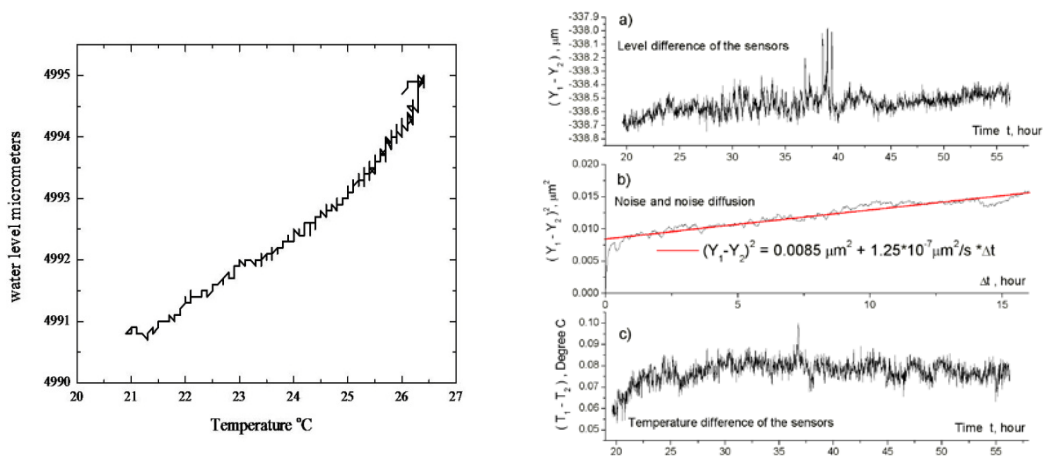

**Figure 7a (left):** Temperature dependence of SASE type sensor. The pool had a fixed disk installed and the sensor was heated and allowed to cool; **Fig.7b (right)** Level difference (a), diffusion (b) and temperature difference (c) of two SAS probes of 2[nd] iteration of design installed 1m apart (April 2001) [11].

# 4. Fermilab Type Sensors

There are 204 superconducting quadrupoles in the Tevatron ring at Fermilab; the quadrupole spacing is 30 meters (97.1 feet). In order to reduce costs instrumentation cost for the 204 superconducting quadrupoles in the Tevatron a simple and inexpensive HLS system was developed. The principle of connected pools of water was used but a commercial proximity sensor was selected to measure the water level. A Baluff model BCAW030NB1Y3 capacitive sensor [10] has the sufficient range and resolution needed for the problem. The sensor is a current source that varies as the capacitance of the system changes. The sensor was adjusted to have a full range of 12.7 mm yielding a 4 micro meter per count resolution. Electronics were developed at Fermilab to readout the sensor and digitize the signals. Figure 8 shows a sensor and electronics.

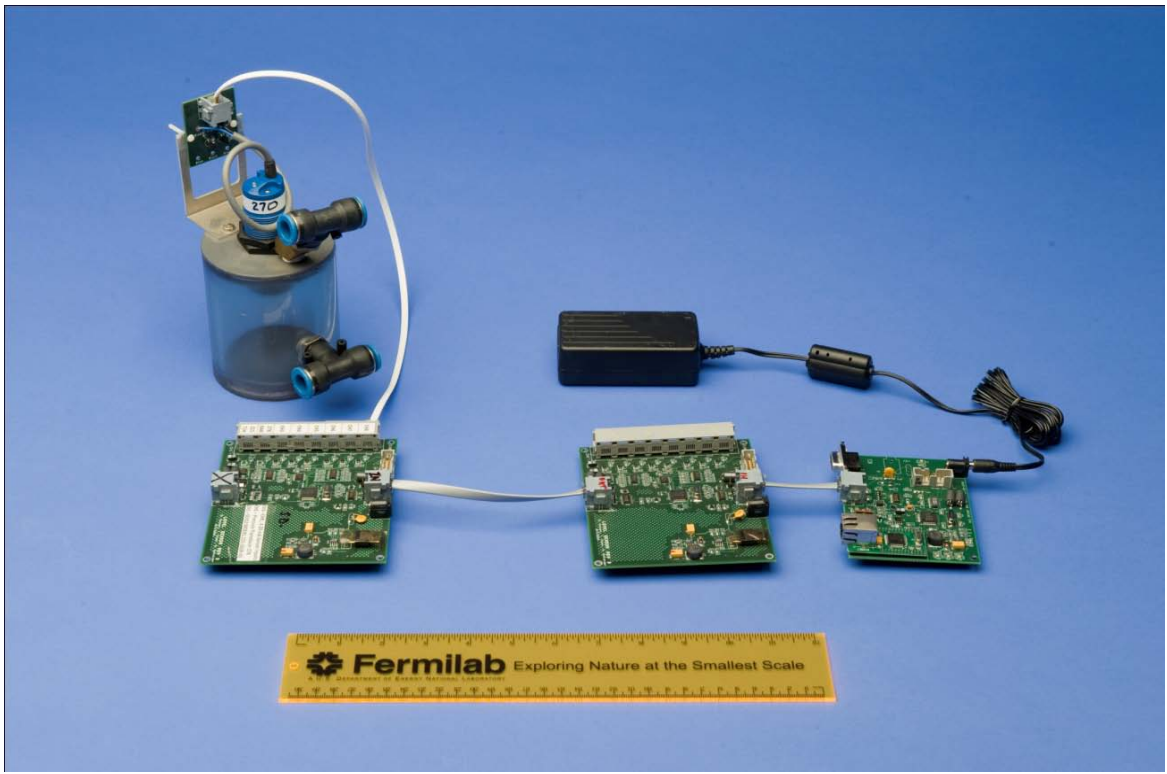

**Figure 8** : The Fermilab style HLS sensor showing the pool and sensor the readout card and driver card along with power supply.

## 4.1 Data Acquisition system

The water level readout system has two control boards consisting of a Driver Board and an 8-Channel Readout Board figure 8 shows the complete system of pool with a sensor and the electronics. The driver board collects data from multiple Readout Boards on a RS485 data link. Each Readout Board has 8 channels of 4-20mA sensor interface inputs. The data from each sensor is collected and stored locally using a micro controller. The sensors are attached to the 8 channel cards using a 4 conductor flat modular cable having RJ11 connectors. Each 4-20ma sensor interface board also has a temperature sensor. The Driver Board has a micro controller with RS232 and Ethernet ports. This allows for data control and display using RS232 or Ethernet. One Ethernet socket is used as a telnet connection and the second supports binary data transfers. The driver board connects to the remote 8 channel Readout Board using a 6 conductor flat modular cable. This cable supplies 24VDC along with differential TX and RX data. Hot swap controller are used on all power connection points.

The TI MSP430 micro controller is used on both cards. The Ethernet connection interface uses a Wiznet NM7010A. On power up firmware in the driver boards' determines how many readout boards are connected and assigns them an incrementing board number. After this power initializing sequence commands can individually access any single board or all boards at once.

### 4.2 Calibration of Tevatron-style HLS

The Balluff sensor has an internal inductor resistor capacitive (LRC) circuit with an adjustable resistor that can be used to set the linear range. Each sensor was placed in a fixture that had an adjustable slide. The sensor was positioned 2 mm above an aluminum block and the resistor was adjusted such that the LED on the body turned from red to green (figure 9). That is the indication that the sensor is within the linear measurement range. The sensor was raised 12.7 mm (1/2 inch) by use of the slide and the resistor was again adjusted to be green. This set the linear range of the circuit to be 12.7 mm. It took several iterations until the proper setting was achieved. The sensor was then positioned close to the block while the readout in milli-volts was recorded. The sensor was moved up in 2.5 mm steps to 12.7 mm then back down the same number of steps while recording the readout. These data were plotted yielding a linear value for the sensor resolution. Those sensors that did not have a linear value were discarded.

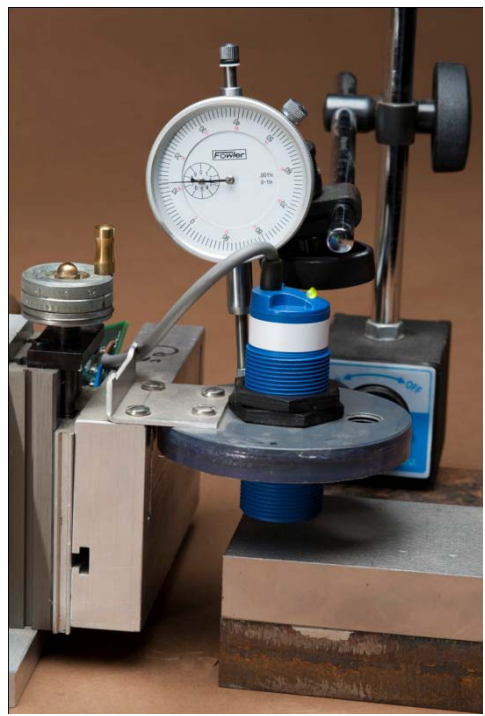

**Figure 9:** Calibration system for Tevatron style HLS, note green LED at top of sensor.

## 5. HLS systems in the Tevatron

### 5.1 Low Beta Quadrupoles

There are two interaction regions in the Tevatron. One is located at the B0 straight section and the other at the D0 straight section. The optics for both regions are the same. It consists of a quadrupole singlet and a quadrupole triplet on either side of the interaction region. To obtain the required high luminosity for the experiments these quadrupoles are superconducting and have high gradient. Motion of these magnets can cause the beam to oscillate around the accelerator and degrade performance.

Each magnet is instrumented with a Budker capacitive HLS sensor of the SAS type described earlier. Each sensor is placed on top of the cryostat in addition there is one sensor mounted on the detector for a total of 9 sensors. There is no leakage of magnetic field from the magnets. A two-pipe fully filled water system was selected to avoid obstacles in the detector hall. In addition for the B0 system the inner two quadrupoles are totally surrounded by shielding steel making access difficult see figure 10. Independent data logging computers are used. Data files from these computers can be obtained for off line analysis. Data are also passed to the Accelerator Division control systems known as ACNET for use by operators and Tevatron system experts.

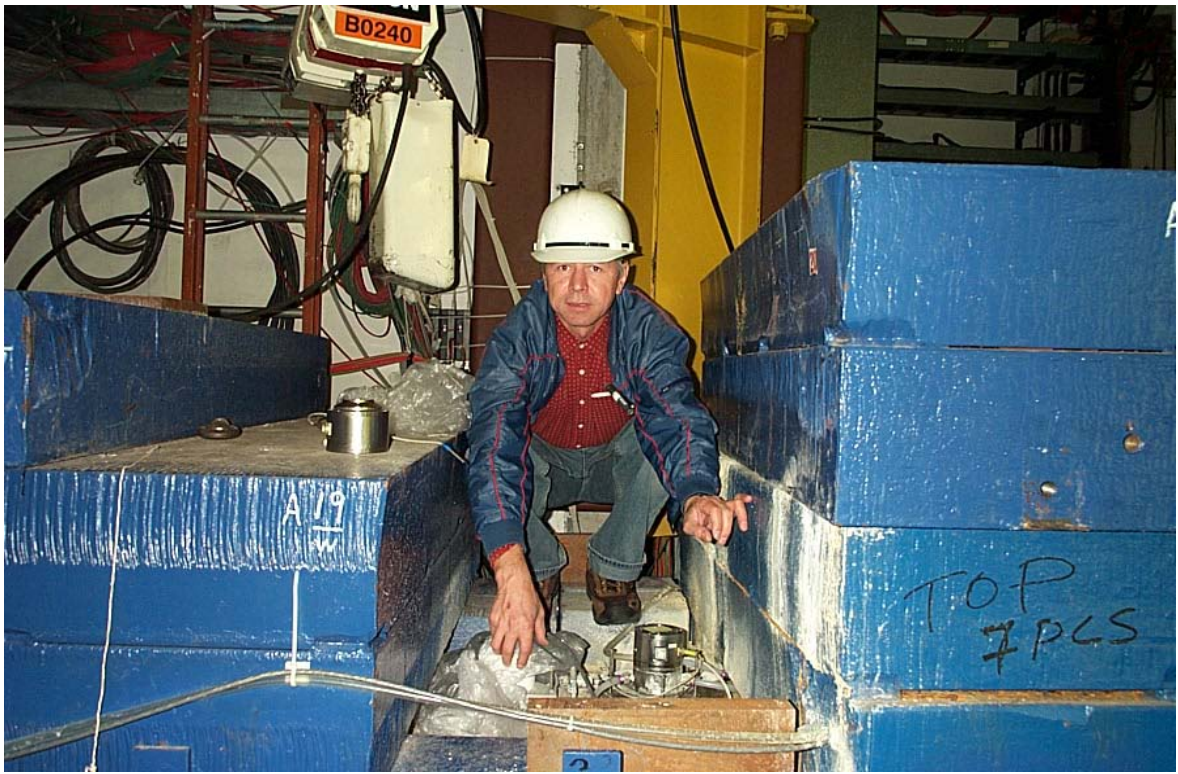

**Figure 10:** Installation of a Budker HLS sensor on top of low beta quadrupole at the B0 interaction region. The steel to the right of will be placed over the top of the sensor.

From time to time and for various reasons such as component failure or beam loss the superconducting magnets in the Tevatron will quench. That is the magnets will rapidly go from 4 Kelvin to a higher temperature and cease to be superconducting. These can be violent events that cause significant magnet motion. Figure 11 show the HLS system on the B side of the B0 detector during a quench and the subsequent recovery from the quench. The BQ4 sensor is closest to the interaction point and shows a 400 micro meter jump in a matter of a few minutes. As cooling down proceeds the magnets and HLS sensors gradually return to similar pre-quench levels. Multiple quenches can cause permanent dislocation of these magnets. Periodic re-alignments of these magnets are required due to the large impact on Tevatron performance.

Another cause of low beta quadrupole motion is temperature changes in the Tevatron tunnel or in the detector halls. Figure 12 shows the effect of the low beta quadrupoles for two weeks in July of 2011. The sensors respond to atmospheric changes in the collision hall. The large spike in the data occurred during an access when 200 tons of steel were moved to gain access to the detector. The magnets did return to their original position after the move.

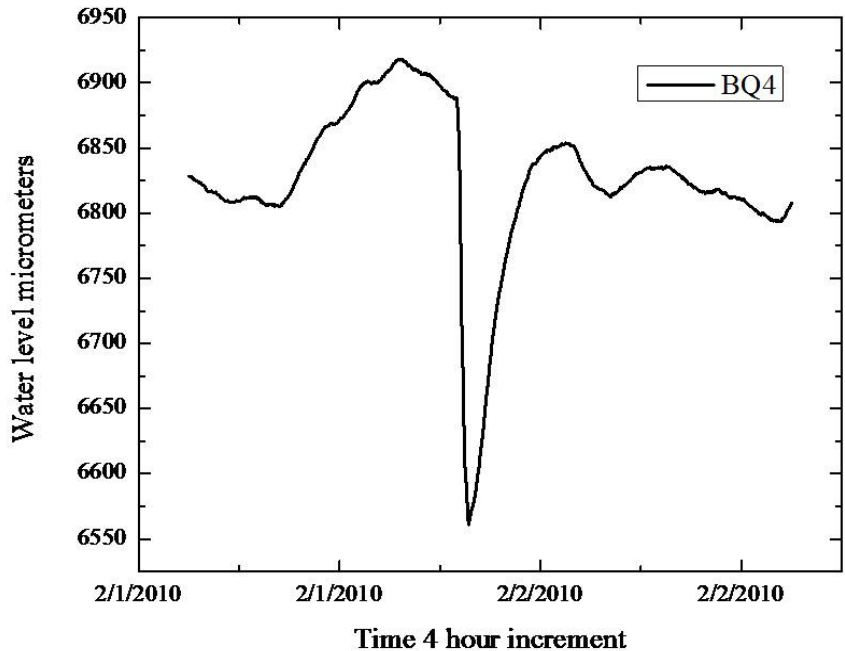

**Figure 11:** Displacement of HLS sensor in micrometers on low beta quadrupole during quench of the magnet.

## 5.2 Regular Superconducting Quadrupoles

The Tevatron is divided into 6 sections (A through F) one kilometer in length. Regular lattice quadrupoles are spaced every 30 meters (97.1 feet) around the ring. The Tevatron has a slight tilt 6 mm over 2 kilometer diameter. To simplify the plumbing and electronics systems were split into 6 sectors corresponding to the sectors in the Tevatron. Each sector is 1 kilometer in length. Sectors A through D have 32 regular lattice quadrupoles and Sectors E and F has 41 lattice quadrupoles. The difference is due to the low beta quadrupoles at the interactions regions at B0 and D0. Between each sensor 12.7 mm (1/2 inch) outer diameter tubing as used to connect the water pools. Groups of 13 or 16 sensors were connected to readout cards with the interface card and power supply located in the above ground service buildings. The read out cards used the TCP/IP port to connect to the Accelerator Controls system (ACNET). Data were available in real time to be plotted or as logged data. Macros were written in Visual Basic for EXCEL that reduced the data and plotted the values versus time.

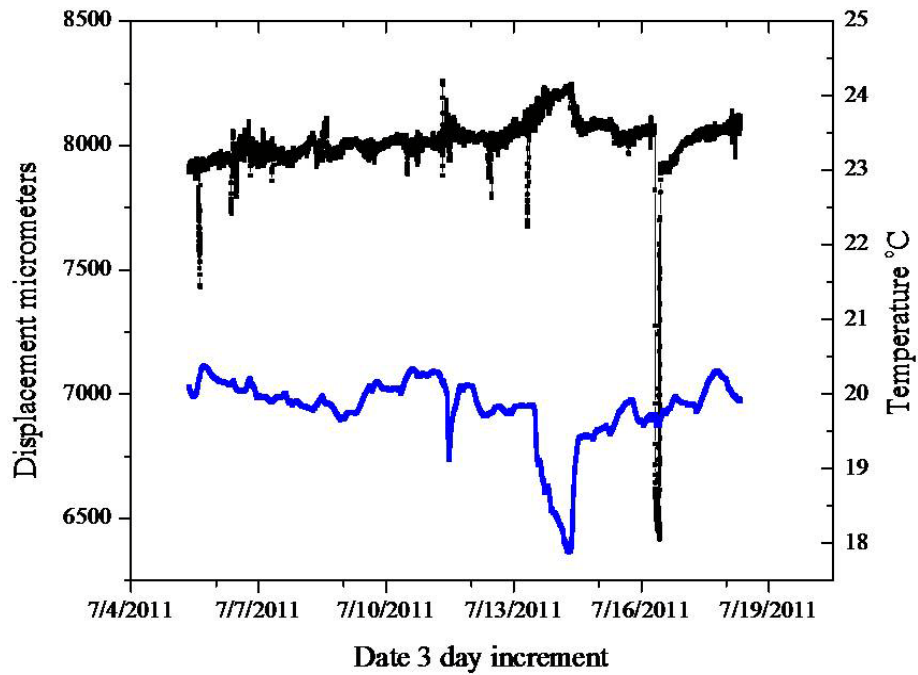

**Figure 12**: Data record from a sensor on low beta quadrupole at the B0 interaction region. The large spike occurred during an access when 200 tons of steel was moved to gain access to detector.

## 6. Deep underground systems

To understand slow ground motion for rock formations where future accelerators maybe built HLS systems were setup in the Galena Platteville dolomite at Fermilab. This formation is typically 100 meters below ground surface in northern Illinois. This is a sedimentary rock common in Illinois. It is mined for use in road construction; there are a number of open pit and underground mines in northern Illinois. The near detector for the MINOS neutrino experiment is sited at the interface between the Maquoketa shale and the Galena Platteville dolomite. Seven kilometers to the south west of Fermilab is the LaFagre Conco mine. This is a room and pillar mine in the Galena Platteville dolomite formation. Both locations were instrumented with HLS systems to study ground motion. Figure 13 shows the Fermilab site and the location of the LaFarge mine in North Aurora Illinois.

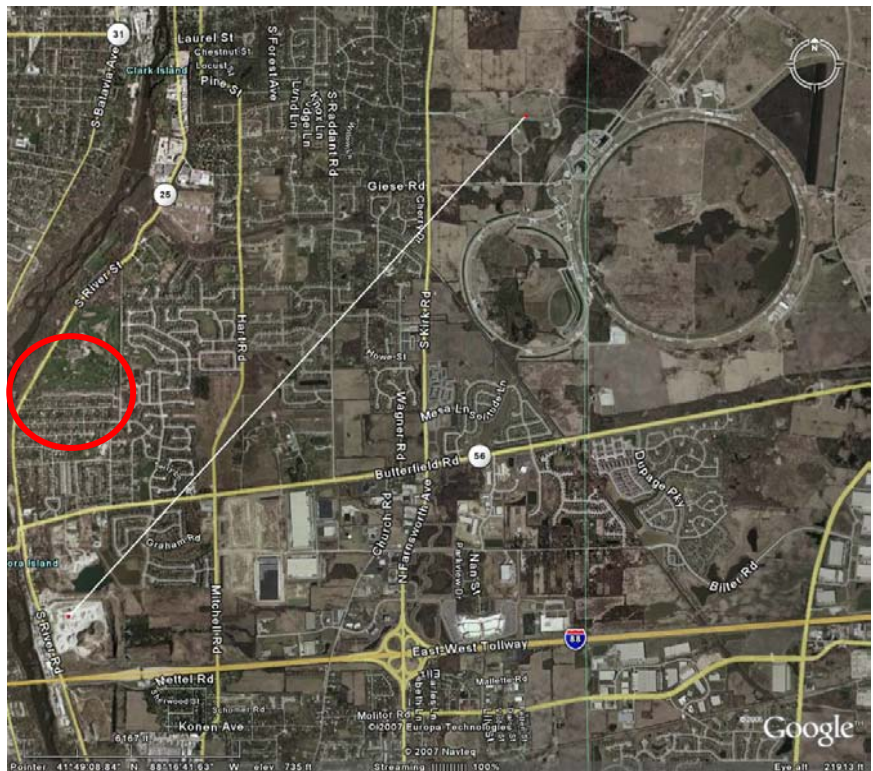

**Figure 13**: Aerial view of Fermilab site showing relation of LaFarge mine in red circle. The distance between the MINOS hall and LaFarge mine is 7 kilometers.

The former Homestake Gold Mine in Lead South Dakota is being prepared for use as a Deep Underground Science and Engineering Laboratory (DUSEL). The mine used a drift and stope method to access the ore deposits, large caverns were then carved out one above the other separated vertically by 50 meters (150 feet). Holes were drilled from the upper to the lower cavern and explosives were used to blast off section of the top of the lower cavern.

There are over 500 kilometers of drifts extending from 100 meters below the surface to more than 2400 meters below the surface. Figure 14 shows a cross section. At the end of mining operations in 2002 the pumps were turned off and water infiltrated the mine flooding from the 2400 meter level (8000 ft) to the 1380 meter (4530 ft) level Dewatering of the mine started in 2008. It is estimated that over 5 giga-tons of water entered the mine and need to be pumped out. This is a significant load and should cause interesting movements in the rock.

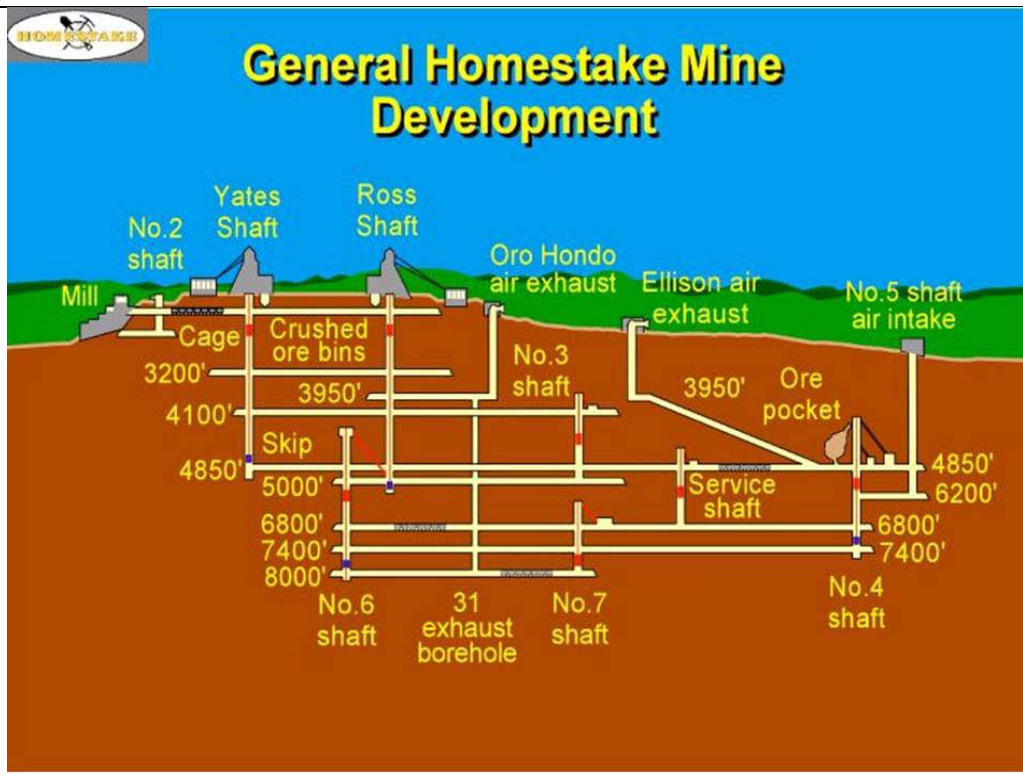

**Figure 14**: Simplified cross section of Homestake mine both HLS installations near the Ross shaft.

### 6.3 MINOS Hall Fermilab site

The MINOS system consists of seven Budker capacitive sensors. Four sensors are arranged in a north south line each 30 meters apart, and three capacitive sensors are arranged on an east west line each 8 meters apart. Initially these sensors were readout using the daisy chain system; in July of 2009 the sensors were switched to Power over the Ethernet. The system uses a two pipe fully filled water connection and the north south water system is separate from the east west system. This is to allow for directional determination of shifts in the floor.

Data acquisition uses a local computer and a NetGear Power over the Ethernet switch. The switch is treated as a private network with the data being logged on the local computer. Remote desktop login is used to access the data files. These files are transferred to another computer for analysis. Once per month the data is posted data base accessible via the Internet. The data consists of air pressure in kilo parcels, the levels in micro meters and the temperature in Celsius.

Tidal motion is clearly visible in the data. The typical 12.4 hour tidal period in addition to the one week period associated with the spring and neap tides. Figure 15 shows typical data for the month of January of 2006. This is the difference in two sensors 90 meter apart. The graph shows the tilt of the floor as a function of time.

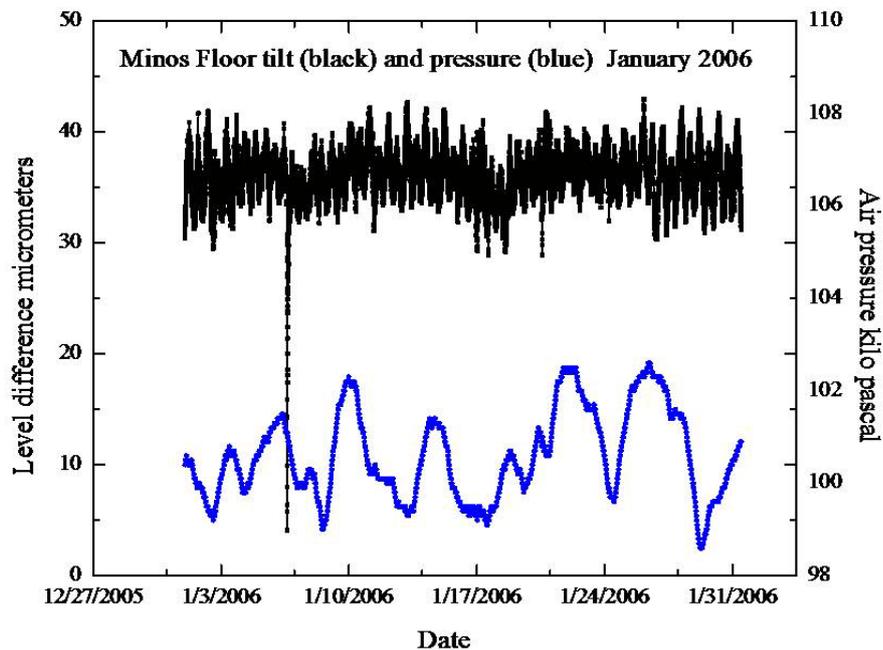

**Figure 15:** Difference in two sensors 90 meters (black) apart January 2006, the 12.4 hour tidal motion and the spring and neap tidal motion are clearly visible. The blue line is air pressure in the mine.

The spike in the data is due to sump pump testing. Ground water infiltrates into the MINOS hall. This water is collected into a sump pit. There are two electric sump pumps that alternate removing water, each pump runs for 10 to 15 minutes alternating every half hour. The sump high limit is set at 75% capacity of the sump and the low stop is set at 25% capacity. Once each month the emergency backup sump pump is tested by running the pump for 30 minutes. This completely drains the sump pit. Figure 16 show a typical testing event. Five minutes after the start of the test the floor of the MINOS hall starts to tilt toward the sump pit and five minutes after the end of the test the floor tilts back. This is due to water flow under the concrete floor of the hall. The floor rebounds to its original slope within one hour.

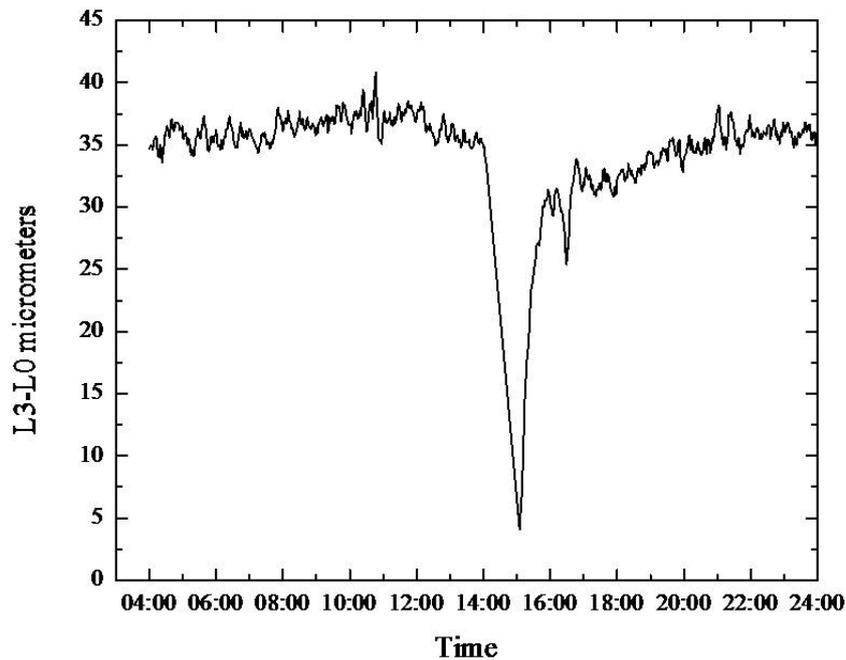

**Figure 16**: Difference in two sensors 90 meters apart during sump pump test in MINOS hall showing the floor tilt due to water motion.

There is also subsidence of the floor. Figure 17 shows 5 years of tilt data taken in the MINOS hall. The graph shows the difference in two sensors 90 meters apart. From the data it can be seen that the floor tilts back and forth over the course of months. There is no discernable pattern to the floor tilt. There is a weak correlation to heavy rain fall events and tilting in the floor. In addition earthquakes can cause floor tilting. In April of 2008 there was a magnitude 5.4 an earthquake in south eastern Illinois on the new Madrid fault (http://earthquake.usgs.gov/ earthquakes/eqinthenews/2008/us2008qza6/#details). The P and S waves from the earthquake were recorded in seismometers located in the MINOS hall. After 28 hours floor tilt was observed in both the north south and east west HLS systems figure 18. The vector sum of the floor tilt pointed in the direction of this earthquake. It is conjectured that either rock or underground water motion was responsible for this subsidence.

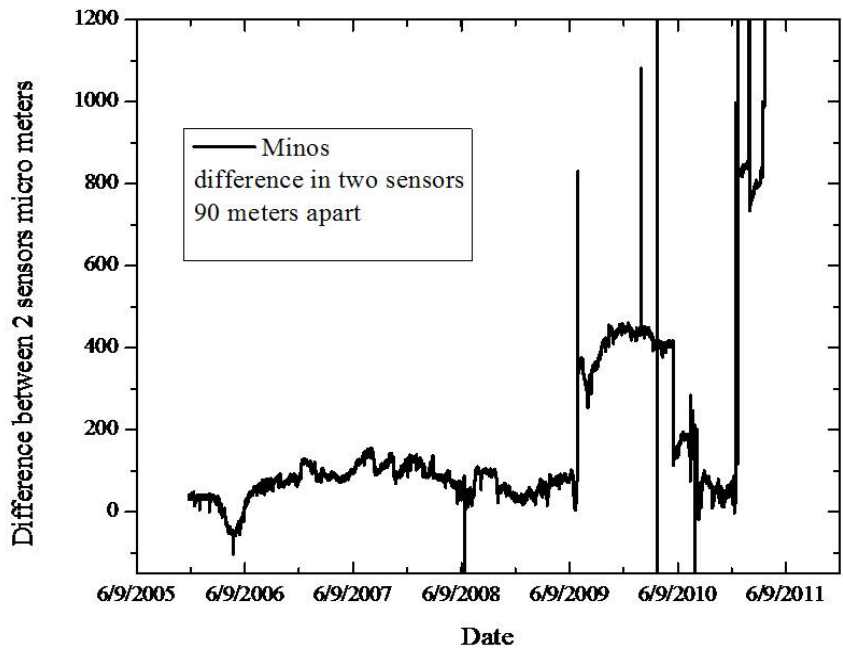

**Figure 17:** Five years of data from MINOS hall at Fermilab this is the difference in two sensor 90 meters apart. The large jumps are due to adding water or sensor change.

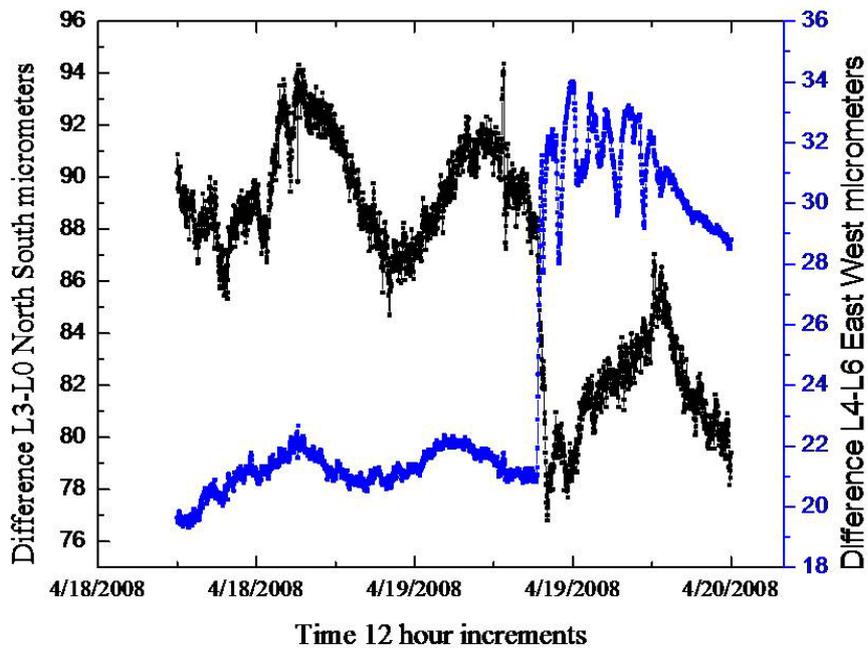

**Figure 18:** Level differences in two sensors for North South and East West arrays in MINOS hall at Fermilab. The sudden change in floor tilt is due to earthquake in southern Illinois 24 hours earlier.

### 6.4 LaFarge Mine North Aurora Illinois

The Lafarge mine is a working dolomite and limestone mine 7 kilometers south west of Fermilab. Several different arrays of HLS systems have been installed since 2000 in this mine. The current working system consists of 4 capacitive sensors with daisy chain read out. Each sensor is 30 meters apart. Concrete pillars were poured on the floor of the drift each was set to a height that allowed for difference in the floor elevation at each sensor station. The first system of 5 sensors each 30 meters apart this system was operational from September 2006 until February of 2008. Due to requirement of the mine this system was moved to an adjacent drift. This system was in operation from June of 2008 until October of 2010. In October of 2010 the system was moved again this time to an east west orientation. The system became operational in December of 2010 and continues taking data.

Figure 19 shows the difference in two sensors for the period June 2008 until October 2010. These data show the tilt of the floor in the mine. The floor showed a continuous tilt for the first year. The direction of the tilt is away the working face of the mine. In November of 2008 the tilting stopped this corresponded to a reduction in mining due to an economic slowdown. When work resumed in the spring of 2009 the floor started tilting again and then stopped when work in the mine stopped in November of 2009. There are also spikes in the floor tilt due to heavy rain events.

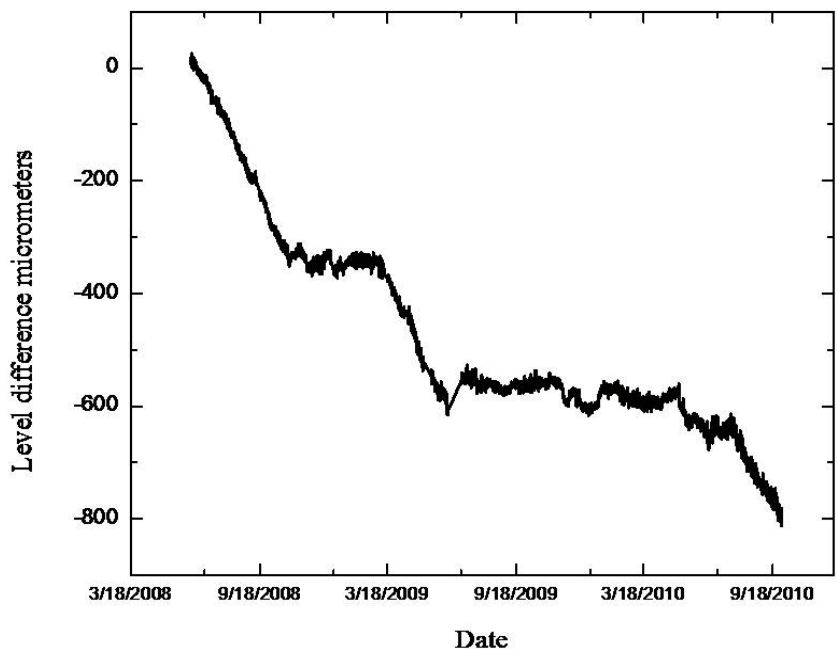

**Figure 19**: Level difference from two sensors LaFarge mine. The mine floor tilts away from the working face of the mine.

This is a working mine using a drill and blast method to extract the rock. Every working day at 14:30 hours local time a blast is set off. These can be clearly seen in figure 20. The spikes correspond to blast and pressure waves caused by blasting in the mine. The weekend time periods show no such spikes.

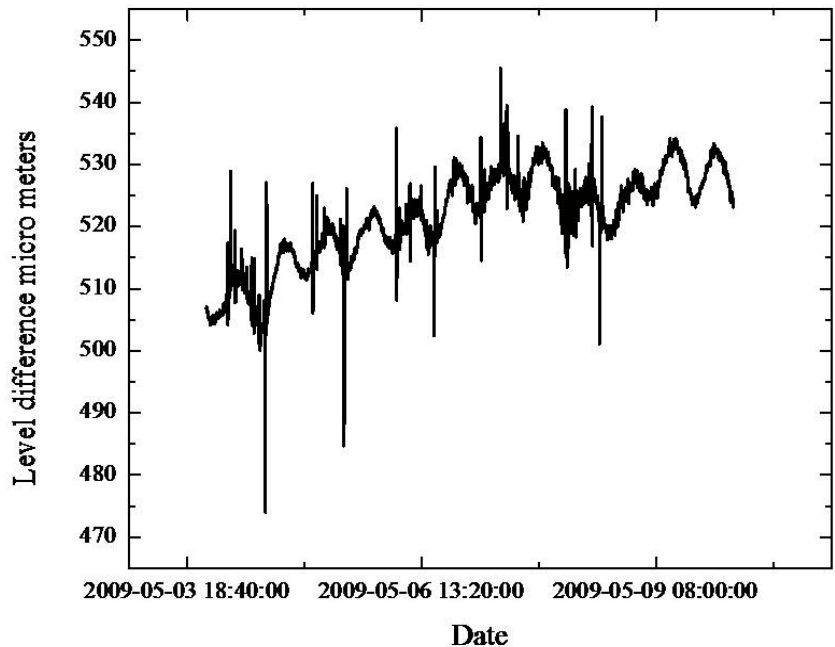

**Figure 20:** Differences in two sensors 120 meters apart LaFarge mine the work week is on the left hand side and show the effect of blasting the weekend is on the right when there is no blasting.

### 6.5 The 2000 foot level at DUSEL

Along a drift connecting the Ross shaft to the stope area three sets of HLS systems were installed see (figure 21). The systems were configured to be at angles with respect to each other to give a vector sum indicating the direction of any motion of the drift. The first two systems used the Fermilab style HLS sensors. These systems were set on concrete pillars poured on the rock bench figure 22. In January of 2009 12 Tevatron style HLS sensors were located in a drift at the 2000 ft level to monitor rock motion during the dewatering process. In February of 2010 6 Budker PoE sensors were installed in the 2000 ft level over the A array of Tevatron sensors. This installation purpose is to test a support stand structure based on the use of rock bolts. A platform made of recycled plastic boards was constructed a 2.4 meter (8 ft) long rock bolt was drilled into the rib of the drift to hold the platform. This is an alternate to pouring concrete bases to hold the sensors.

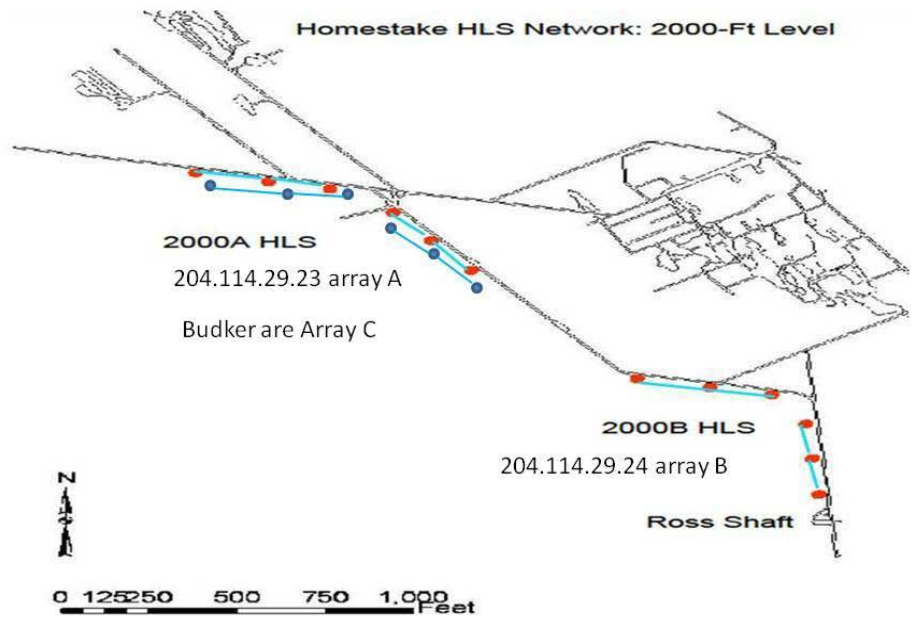

**Figure 21**: Layout of the 2000 foot level Homestake mine showing locations of the three HLS arrays. Note the separation of Array C is half that of Array A.

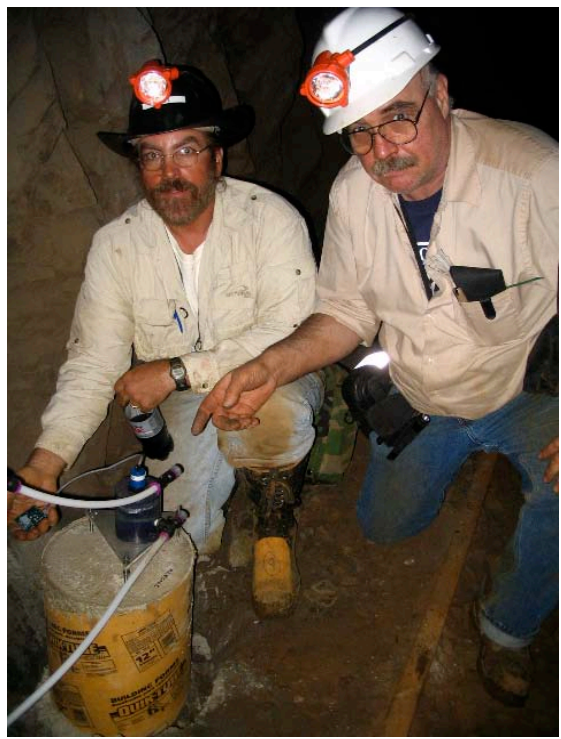

**Figure 22:** Two of the authors (L Stetler and J Volk) near the Tevatron style HLS sensor 2000 foot level Homestake mine.

### 6.6 The 4850 foot level at DUSEL

In June of 2011, 12 Budker Ultra Sonic sensors were installed at the 4850 foot level in two separate plumbing systems figure 23. The sensors were placed on the concrete floor. Differences in floor elevation from one sensor to another were made up by the use of granite slabs obtained from counter top manufactures. These slabs are the waste cut outs. The system is a single half filled pipe. Again two separate plumbing systems were used these were set at an angle of 25 degrees so as to obtain the vector sum of any floor motion. The sensors were spaced 7 meters apart. This was driven by the available length of the drift. Sensors number 5 and 6 the terminal sensors of each plumbing system were set within 1.5 meters of each other (figure 24). This configuration allows for comparison of the system noise. Figure 25 shows typical data for this system the difference in two sensors 30 meters apart. The tidal motion can be seen in the data. Figure 26 shows the Fast Fourier Transform of the data presented in figure 25. The magnitude of the spectrum shows a peak at 12.4 hours typical of tidal motion.

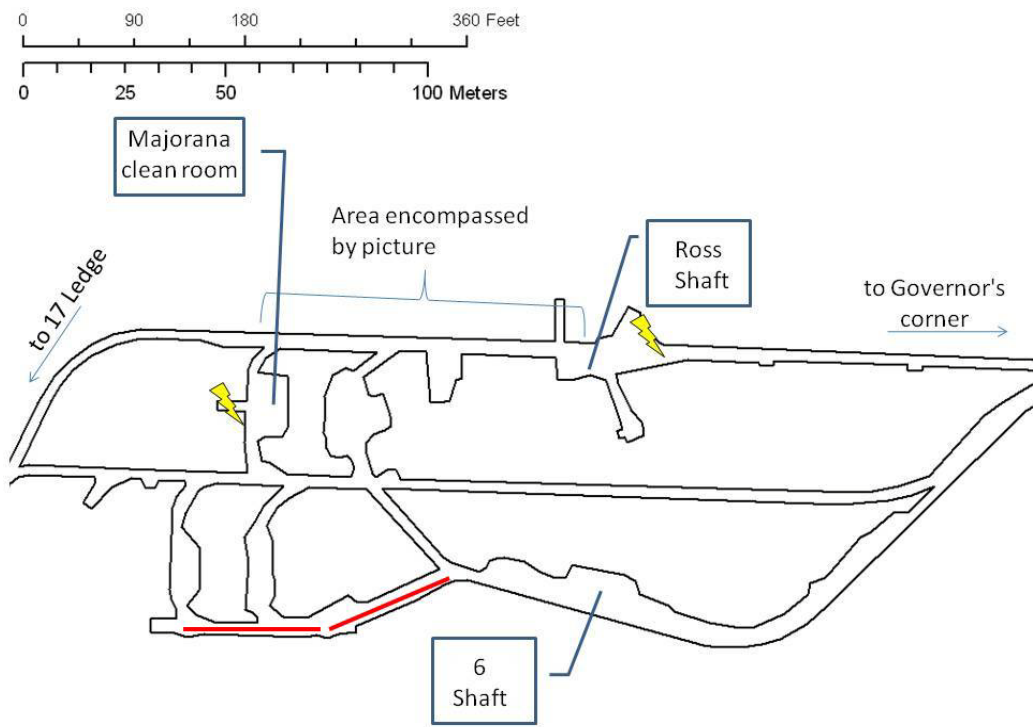

**Figure 23**: Layout of the 4850 foot level near the Ross shaft the red lines indicate the location of the HLS array.

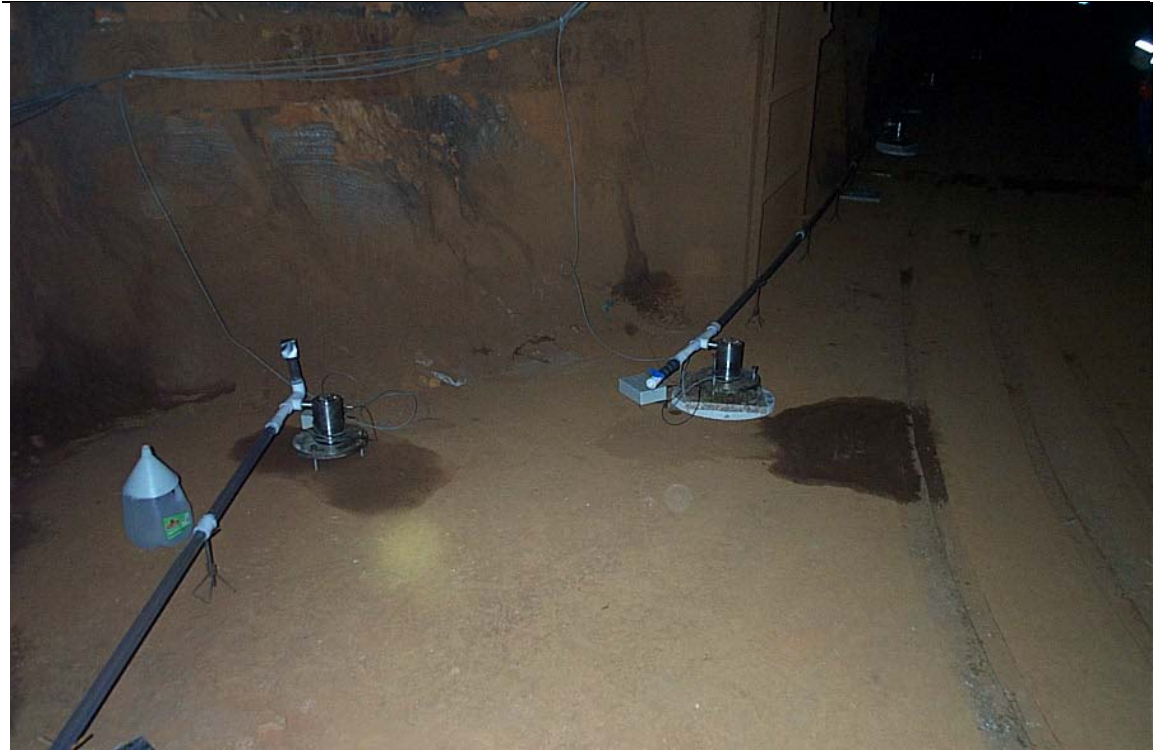

**Figure 24:** Sensors L5 and L6 at the 4850 foot level Homestake mine, note granite slabs used to adjust height.

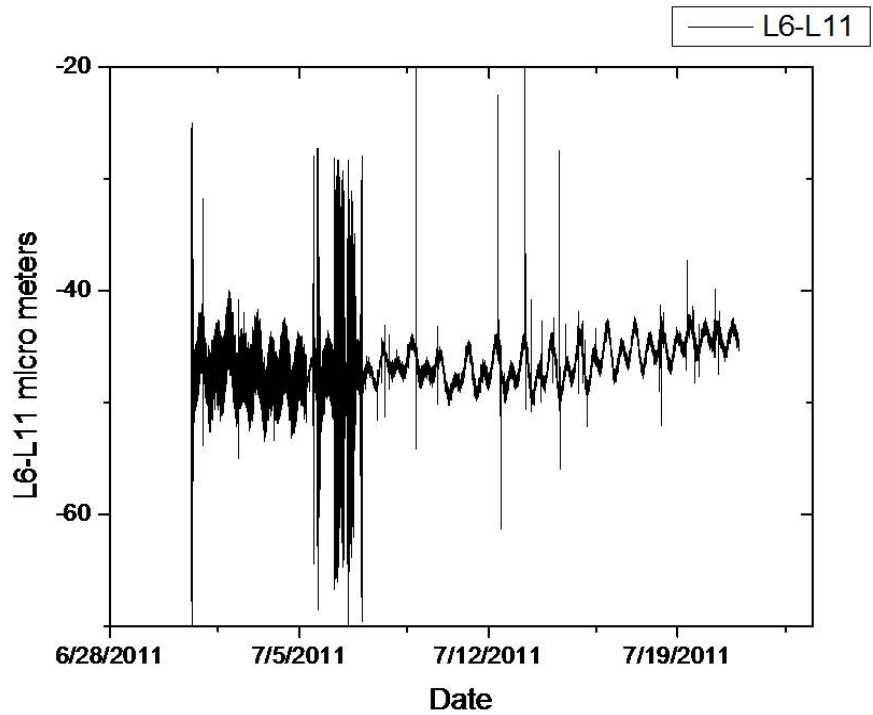

**Figure 25:** Difference in two sensors 29.5 meters apart at 4850 foot level of the Homestake mine. Tidal movement is clearly visible. The motion in the first part of the graph is not as of yet understood.

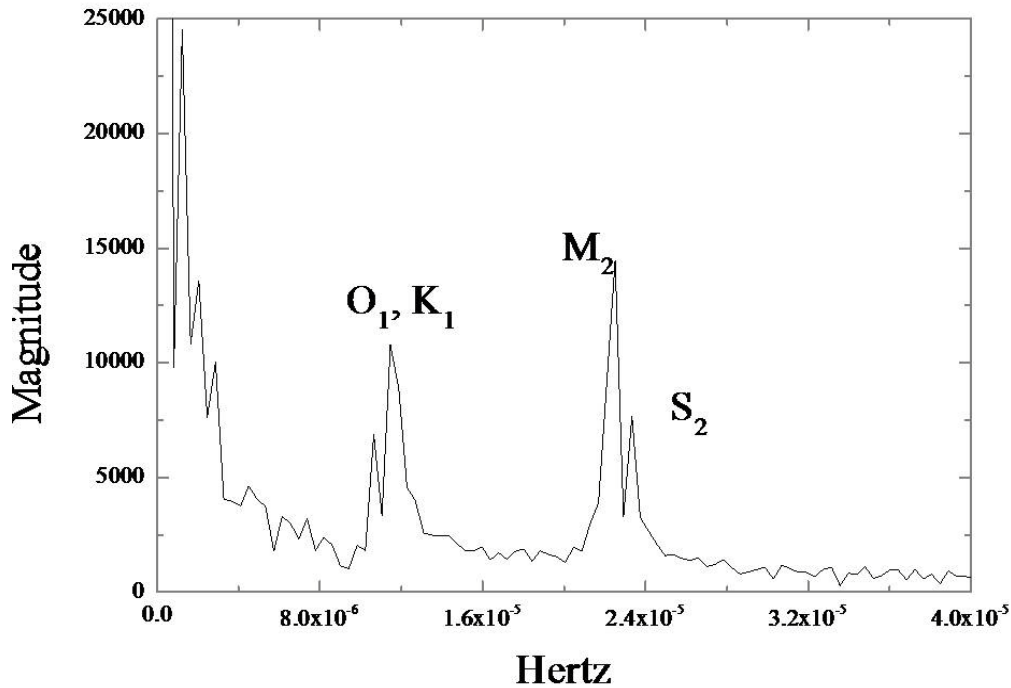

**Figure 26**: Fast Fourier transform of July data from 4850 foot level showing tidal peak at $2.3 \times 10^{-5}$ hertz. $M_2$ is principle lunar and $S_2$ is principle solar tides $O_1$ is a secondary lunar and $K_1$ is secondary lunar and solar tide.

It is anticipated that soon this array will move south and west to an area known as the 17 Ledge Motor Barn. This is a larger area far from Ross shaft, some 600 meters, and will allow for the arrays to be positioned at right angles with respect to each other. That will allow for determining vector sums of any floor tilting as a means to determine location and cause.

## 7. Summary and conclusions

Fermilab and the Budker Institute have developed and deployed several HLS system to study slow ground motion. Natural sources of ground motion such as tidal motion, subsidence caused by earth quakes and pumping of ground water have been observed. In addition motion caused by cultural effects such as quenches of cryogenic magnets, blasting and traffic have been observed. Similar systems will be needed in the next generation of energy frontier accelerators to provide information to stabilize the beams. The HLS systems have been widely used for exploration of various natural geophysical phenomena at accelerators, such as ground diffusion governed by so called "ATL law" [12].

## Acknowledgments

The authors wish to thank the technical staff of both Fermi National Accelerator Laboratory and the Budker Institute of Nuclear Physics for assistance in developing the electronics and software for the various HLS sensors. We also wish to thank the staff at the Sanford Laboratory for assistance with the installation of systems in the Homestake mine in Lead South Dakota, in particular Tom Trancynger. We also wish to thank Brian Washkowika of the Conco Lafarge Mine in North Aurora Illinois for assistance with installation of the HLS sensors at that facility. Authors are very thankful to Drs R.Ruland and A.Seryi of SLAC for many years of mutually fruitful collaboration in developing high accuracy HLS for accelerators. We also wish to thank Andreas Herty of CERN for comments on the manuscript.